\documentstyle[12pt]{article}
\newcommand{\norm}[1]{{\protect\normalsize{#1}}}
\newcommand{\LAP}
{{\small E}\norm{N}{\large S}{\Large L}{\large A}\norm{P}{\small P}}

\newcommand{\be}{\begin{equation}}
\newcommand{\ee}{\end{equation}}
\newcommand{\bea}{\begin{eqnarray}}
\newcommand{\ena}{\end{eqnarray}}
\newcommand{\beano}{\begin{eqnarray*}}
\newcommand{\enano}{\end{eqnarray*}}
\newcommand{\sect}[1]{\setcounter{equation}{0}\section{#1}}
\newcommand{\vs}[1]{\rule[- #1 mm]{0mm}{#1 mm}}

\newcommand{\sm}[2]{\frac{\mbox{\footnotesize #1}\vs{-2}}
                   {\vs{-2}\mbox{\footnotesize #2}}}

\newcommand{\shalf}{\sm{1}{2}}

\newcommand{\Stop}{s^{top}}
\newcommand{\ca}{\mbox{$\cal{A}$}}

\newcommand{\cd}{\mbox{$\cal{D}$}}
\newcommand{{\cg}}{\mbox{$\cal{G}$}}

\newcommand{\cw}{\mbox{$\cal{W}$}}
\newcommand{\m}{\mbox{\large{$m$}}}
\newcommand{\e}{\mbox{\large{$e$}}}

\newcommand{\om}{\omega}
\newcommand{\Om}{\Omega}
\newcommand{\tildom}{\widetilde{\omega}}
\newcommand{\Tildom}{\widetilde{\Omega}}

\newcommand{\cj}{\mbox{${\cal I}$}}

\newcommand{\under}{\underline}

\def\annexe#1#2{\def\thesection{\Alph{section}}\section*{#2}
                \setcounter{section}{#1}  }

\begin{document}

\newpage
\pagestyle{empty}
\setcounter{page}{0}

\def\logolapin{
  \raisebox{-1.2cm}{\epsfbox{/lapphp8/keklapp/ragoucy/paper/enslapp.ps}}}
\def\logolight{{\bf {\large E}{\Large N}{\LARGE S}{\huge L}{\LARGE
        A}{\Large P}{\large P} }}
\def\logoenslapp{\logolight}
%
%
%
\hbox to \hsize{
\hss
\begin{minipage}{5.2cm}
  \begin{center}
    {\bf Groupe d'Annecy\\ \ \\
      Laboratoire d'Annecy-le-Vieux de Physique des Particules}
  \end{center}
\end{minipage}
\hfill
\logoenslapp
\hfill
\begin{minipage}{4.2cm}
  \begin{center}
    {\bf Groupe de Lyon\\ \ \\
      {\'E}cole Normale Sup{\'e}rieure de Lyon}
  \end{center}
\end{minipage}
\hss}

\vspace {.3cm}
\centerline{\rule{12cm}{.42mm}}

\vs{5}

\begin{center}

{\bf DE LA FIXATION DE JAUGE CONSIDEREE COMME UN DES BEAUX ARTS 
ET DE LA SYMETRIE DE SLAVNOV QUI S'ENSUIT}\\[1cm]

\vs{10}

{\large R. Stora}\\[1cm]

{\em Laboratoire de Physique Th{\'e}orique }\LAP\footnote{URA 14-36
du CNRS, associ{\'e}e {\`a} l'Ecole Normale Sup{\'e}rieure de Lyon et {\`a}
l'Universit{\'e} de Savoie}\\
{\em LAPP, Chemin de Bellevue, BP 110\\ 
F-74941 Annecy-le-Vieux Cedex, France.\\[0.2cm]
and\\[0.2cm]
CERN, TH-Division, 1211 Gen{\`e}ve 23, Switzerland}\\

\end{center}
\vs{15}

\centerline{ {\bf Abstract}}

\indent

La fixation de jauge est d{\'e}finie comme l'op{\'e}ration permettant
d'exprimer une int{\'e}grale sur un espace d'orbite comme int{\'e}grale
sur le fibr{\'e} principal correspondant. Quand la fibre est non
compacte cette op{\'e}ration met en jeu une classe de cohomologie {\`a}
support compact -ou {\`a} d{\'e}croissance rapide- de celle-ci. La
sym{\'e}trie de Slavnov est l'expression alg{\'e}brique de
l'ambiguit{\'e} de cette construction.

\vs{5}

\rightline{\LAP-A-620/96}
\rightline{October 1996}

\newpage
\pagestyle{plain}

\section{Int{\'e}gration de certaines classes\\de cohomologie
{\'e}quivariantes\protect\footnotemark[1]
\protect\footnotetext[1]{Les notions relatives {\`a}
l'{\'e}quivariance sont rappel{\'e}es dans l'appendice A} }

\indent

La fixation de jauge est l'un des d{\'e}tails techniques in{\'e}vitables
qui embarrasse le paysage des th{\'e}ories de jauge. Sa n{\'e}cessit{\'e}
est d{\^u}e {\`a} la non compacit{\'e} du groupe de jauge. La situation
est la suivante : soit $\ca$ un fibr{\'e} principal de fibre $\cg$, un
groupe de Lie connexe non compact. On supposera $\ca$ non trivial,
en g{\'e}n{\'e}ral. Par exemple, $\ca$ sera l'espace des connexions
principales $\under{a}$ sur un fibr{\'e} principal $P(M,G)$ o{\`u} $M$
est une vari{\'e}t{\'e} compacte, $G$ un groupe de Lie compact, et o{\`u}
$\cg$ est le groupe de jauge de ce fibr{\'e} d{\'e}fini de sorte que
$\ca$ soit un fibr{\'e} principal. Les difficult{\'e}s li{\'e}es {\`a} la
dimension infinie sont rappel{\'e}es dans l'appendice B. 

Soit
$\Theta$ un repr{\'e}sentatif d'une classe de cohomologie
{\'e}quivariante de $\ca$ de dimension $| \ca / \cg |$ et
$\widetilde{\Theta}$ la forme basique obtenue en choississant une
connexion $\widetilde{\om}$ sur $\ca$, de courbure $\widetilde{\Omega}$.
De fa{\c c}on g{\'e}n{\'e}rale, soit $\widetilde{\Theta}$ une forme basique
sur $\ca$, de dimension $| \ca / \cg |$. On cherche une repr{\'e}sentation
int{\'e}grale sur $\ca$ de l'int{\'e}grale $\int_{\ca / \cg}
\widetilde{\Theta}$, o{\`u}, par abus de notation, on a confondu
$\widetilde{\Theta}$ avec la forme qu'elle d{\'e}finit sur $\ca / \cg$.
Dans le cas des th{\'e}ories de jauge, on part d'une forme  $\Theta$
$\cg$ invariante sur $\ca$, de degr{\'e} maximum et on construit la
forme de Ruelle Sullivan $\widetilde{\Theta}_{RS}$ \cite{1} associ{\'e}e
{\`a} une forme volume invariante sur Lie $\cg$, d{\'e}finie {\`a} un
scalaire multiplicatif positif pr{\`e}s (cf. paragraphe 2). Pour
exprimer l'int{\'e}grale sur $\ca / \cg$, on peut choisir, en
supposant $\ca / \cg$ paracompacte un recouvrement $\{ U_i, i \in I
\}$ localement fini et une partition de l'unit{\'e} $\{ \theta_i
(\dot{a}_i), i \in I \}$ o{\`u} $\{ \dot{a}_i \}$ d{\'e}signe un choix
de coordonn{\'e}es dans l'ouvert $U_i$, ainsi que des sections locales
$\sigma_i$ au dessus des $U_i$, repr{\'e}sent{\'e}es par des
{\'e}quations locales $g_i (a_i) =0$ o{\`u} les $a_i$ sont des coordonn{\'e}es locales
de $\ca$ au dessus de $U_i$. On peut alors {\'e}crire
\be
\int_{\ca / \cg} \widetilde{\Theta} = \sum_{i \in I} \int_{\ca / \cg}
\theta_i (\dot{a}) \widetilde{\Theta}) \int_{fibre} \delta (g_i) (
\wedge \delta g_i)
\ee
o{\`u} on a ins{\'e}r{\'e} l'int{\'e}grale sur la fibre d'un repr{\'e}sentatif
du dual de Poincar{\'e} de l'image $\sum_i$ de $\sigma_i$, au dessus
de $U_i$. $\delta$ d{\'e}note la diff{\'e}rentielle sur $\ca$. Les $g_i$
sont des fonctions {\`a} valeur dans un espace vectoriel de dimension
$|\cg|$ {\'e}gale {\`a} la dimension de $\cg$, et $\delta (g_i) (\wedge
\delta g_i)$ est par construction ind{\'e}pendant du choix d'une base
dans cet espace.

Comme $\widetilde{\Theta}$ est de degr{\'e} maximum, on peut restreindre
$\wedge \delta g_i$ {\`a} la fibre en utilisant une connexion
$\widetilde{\om}$ arbitraire sur $\ca$ en {\'e}crivant
\be
\delta g_i = \frac{\delta g_i}{\delta a_i} \delta a_i = \frac{\delta
g_i}{\delta a_i} ( \widetilde{\psi}_i + \ell (\widetilde{\om}) a_i)
\ee
o{\`u} $\widetilde{\psi}_i$ est la partie horizontale de $\delta a_i$ pour
la connexion $\widetilde{\om}$, $\ell(\widetilde{\om})$ d{\'e}signe la
d{\'e}riv{\'e}e de Lie le long du champ de vecteur fondamental
correspondant {\`a} l'{\'e}l{\'e}ment $\widetilde{\om}$ de Lie $\cg$.
La restriction {\`a} la fibre consiste {\`a} oublier $\widetilde{\psi}_i
$, autrement dit {\`a} travailler modulo l'id{\'e}al diff{\'e}rentiel
$\cj^+_h$ engendr{\'e} par les formes horizontales. Du m{\^e}me coup
le choix de $\tildom$ n'importe pas, la diff{\'e}rence de deux
connexions {\'e}tant horizontale.
Introduisant une "fonction $\delta$ fermionique" au moyen d'une
int{\'e}gration de Berezin, on peut r{\'e}crire 
\be
\int_{\ca / \cg} \widetilde{\Theta} = \int_a \int \cd \om
\widetilde{\Theta} (\wedge \widetilde{\om}) \gamma(a,\om)
\ee
avec
\be
\gamma (a, \om) = \sum_i \theta_i (\dot{a}) \delta(g_i) \wedge \m_i
\om
\ee
o{\`u}
\be
\m_i \om = \frac{\delta g_i}{ \delta a_i} \ell (\om) a_i
\ee
$\gamma (a, \om)$ repr{\'e}sente la projection sur la fibre de la
forme
\be
\gamma (a, \widetilde{\om}) = \sum_i \theta_i (\dot{a}) \delta (g_i)
\wedge m_i \widetilde{\om}
\ee
ind{\'e}pendamment du choix de $\widetilde{\om}$. 

Les volumes utilis{\'e}s dans $\cd \om$ et dans $\wedge
\widetilde{\om}$ sont duaux l'un de l'autre.

\indent

Par construction, $\gamma (a, \widetilde{\om})$ est de degr{\'e} $|\cg |$
puisque, les sections $\Sigma_i$ {\'e}tant transverses aux fibres, les
op{\'e}rateurs $\m_i$ sont inversibles. De plus, ind{\'e}pendamment de
la fibre 
\be
\int_{fibre} \gamma (a, \widetilde{\om}) =1
\ee

On peut prendre $\gamma$ {\`a} support compact ou {\`a}
d{\'e}croissance rapide le long des fibres.

Il est clair que, dans le calcul pr{\'e}c{\'e}dent, on aurait pu
remplacer $\gamma$ par une forme de (fixation de) jauge avec les
propri{\'e}t{\'e}s suivantes:
$\gamma$ repr{\'e}sente la projection sur la fibre d'une forme de
degr{\'e} $| \cg |$ d'int{\'e}grale sur la fibre {\'e}gale {\`a} 1 quelque
soit la fibre. Il s'ensuit que si $s$ d{\'e}note la diff{\'e}rentielle
induite par projection sur la fibre $(\Omega^* (\ca) \rightarrow
\Om^* (\ca) / \cj^+_h)$, o{\`u} $\cj^+_h$ est l'id{\'e}al engendr{\'e}
par les formes horizontales de degr{\'e} strictement positif)
\bea
s \om &=& - \shalf [\om, \om ] \nonumber \\
s a &=& \ell(\om) a
\ena
on reconna{\^\i}t ici la partie g{\'e}om{\'e}trique de la sym{\'e}trie de
Slavnov\cite{2} qui, ainsi qu'on le voit est g{\'e}om{\'e}triquement
naturelle. 

La diff{\'e}rentielle $s$ est bien la projection sur la fibre de
l'op{\'e}ration $\Stop$ d{\'e}riv{\'e}e de la diff{\'e}rentielle de
l'alg{\`e}bre de Weyl et de celle li{\'e}e {\`a} l'action de $\cg$ sur
$\ca$ (cf. Appendice A)
\be
\begin{array}{ll}
\Stop \widetilde{\om} = \widetilde{\Om} - \shalf [ \tildom ,
\tildom ] & \Stop a = \widetilde{\psi} + \ell(\tildom) a \\
\Stop \Tildom = - [ \tildom, \Tildom ] & \Stop \widetilde{\psi} = -
\ell(\Tildom) a - \ell(\tildom) \widetilde{\psi}
\end{array}
\ee
puisque $\widetilde{\psi}, \Tildom, [\tildom, \Tildom],
\ell(\tildom) \widetilde{\psi}$ appartiennent {\`a} $\cj^+_h$.

Comme $\gamma$ est de degr{\'e} maximum on a trivialement
\be
s \gamma (a, \om) = 0
\ee
De plus, comme $\gamma$ repr{\'e}sente une classe de cohomologie {\`a}
support compact (ou {\`a} d{\'e}croissance rapide) de $\cg$ (qu'on a
suppos{\'e} connexe), $\gamma$ est d{\'e}fini {\`a} un cobord pr{\`e}s :
\be
\gamma \rightarrow \gamma + s \chi
\ee
o{\`u} $\chi$ est la projection sur la fibre d'une forme {\`a} support
compact ou {\`a} d{\'e}croissance rapide.

Si $\cg$ {\'e}tait compact, on pourrait choisir $\gamma =1$ (la jauge
unitaire des physiciens), mais dans le cas non compact ce choix est
impossible.

\vs{2}

{\bf Remarque}

La construction ci-dessus, qui pr{\'e}conise l'int{\'e}gration sur la
fibre comme elle est esquiss{\'e}e dans le livre de J.~Zinn
Justin\cite{3}, fournit une alternative {\`a} la m{\'e}thode initiale de
Faddeev et Popov\cite{4} bas{\'e}e sur la factorisation du volume du
groupe de jauge qui n'est pas adapt{\'e}e au cas d'un groupe $\cg$ non
compact, m{\^e}me en dimension finie.

L'espace des formes de jauge est non vide, par construction,
convexe. Reste {\`a} construire des formes de jauge "commodes" et
respectant la g{\'e}om{\'e}trie. A ce point, faute de mieux, on
conjecture l'existence de formes de jauge du type
\be
\int \cd \bar{\om} \cd b \; \e^{s(\bar{\om} g(a, \dot{a}) + i
\bar{\om} \varphi (b, \dot{a}))}
\ee
o{\`u} l'op{\'e}ration $s$ est {\'e}tendue aux variables d'int{\'e}gration
selon
\bea
s \bar{\om} &=& ib \nonumber \\
s b &=& 0,
\ena
o{\`u} $g(a, \dot{a})$ est une fonction $\ca$ jauge telle que $\m(a,
\dot{a})$ est partout inversible, et o{\`u} $b \varphi (b, \dot{a})$
est positive croissante {\`a} l'infini. Si cette classe de formes de
jauge est non vide, elle conduit vraisemblablement {\`a} des choix de
jauge non renormalisables et ou non locaux, {\`a} moins qu'on ne
r{\'e}ussisse {\`a} simuler les effets non locaux au moyen de
l'introduction de champs locaux auxiliaires.

\sect{Le cas des th{\'e}ories de jauge}

\indent

On se donne sur $\ca$ une forme $\Theta$ $\cg$ invariante de degr{\'e}
maximum. Si on se donne une forme volume $\mu$ invariante sur $\cg$,
il lui correspond une forme volume duale sur Lie $\cg \; 
\widetilde{\mu}$. On d{\'e}finit
\be
\Theta_{RS} = i(\widetilde{\mu}) \Theta
\ee
o{\`u} le symbole de contraction $i(\widetilde{\mu})$ est d{\'e}fini en
repr{\'e}sentant Lie $\cg$ par les champs de vecteur fondamentaux
appropri{\'e}s. L'invariance de $\Theta$ assure que $\Theta_{RS}$ est
ferm{\'e}e, et, $\Theta_{RS}$ est horizontale, par construction. On se
ram{\`e}ne ainsi au probl{\`e}me d{\'e}crit dans le paragraphe 1 avec
\be
\widetilde{\Theta}_{RS} \wedge \widetilde{\om} = \Theta .
\ee

\setcounter{equation}{0}

\annexe{1}{Appendice A}

\indent

On rappelle la terminologie classique relative {\`a} la cohomologie
{\'e}quivariante.

Soit $M$ une vari{\'e}t{\'e} sur laquelle op{\`e}re un groupe de Lie
connexe $\cg$ d'alg{\`e}bre de Lie Lie $\cg$ repr{\'e}sent{\'e}e par des
champs de vecteurs sur $M$ :
\be
\lambda \in Lie \; \cg \rightarrow \under{\lambda} \in Vect \;  M
\ee
tels que le crochet de Lie $\cg$ soit repr{\'e}sent{\'e} par le crochet
de Lie de Vect $M$.

Soit $\Omega^* (M)$ l'alg{\`e}bre des formes ext{\'e}rieures sur $M$
munie de la diff{\'e}rentielle $d_M$.

On d{\'e}finit sur $\Omega^*(M)$
\be
i_M (\lambda) = i_M (\under{\lambda})
\ee
le produit int{\'e}rieur par $\under{\lambda} \in \; Vect \; M$
\be
\ell_M (\lambda) = \ell_M (\under{\lambda}) = [i_M
(\under{\lambda}), d_M ]
\ee
la d{\'e}riv{\'e}e de Lie le long de $\lambda$.

Les formes horizontales $\om_h$ sont celles pour lesquelles
\be
i_M (\lambda) \om_h =0 \; \; \; \; \forall \lambda \in Lie \; \cg
\ee

Les formes invariantes $\om_{inv}$ sont celles pour lesquelles
\be
\ell_M (\lambda) \om_{inv} =0 \; \; \; \; \forall \lambda \in Lie
\; \cg \ee

Les formes qui sont {\`a} la fois horizontales et invariantes sont
appel{\'e}es basiques. La cohomologie basique de $M$ est la
cohomologie des formes basiques pour la diff{\'e}rentielle $d_M$. Ces notions se
g{\'e}n{\'e}ralisent {\`a} toute alg{\`e}bre diff{\'e}rentielle gradu{\'e}e
commutative $(E, d_E)$ avec une action de Lie $\cg : i_E (\lambda)$,
d{\'e}rivation gradu{\'e}e de degr{\'e} $-1, \ell_E (\lambda) = [i_E
(\lambda), d_E ]_+$ telles que
\bea
{[} i_E (\lambda), i_E(\lambda ') ] &=& 0 \nonumber \\
{[} \ell_E (\lambda), i_E (\lambda ') ] &=& i_E ([\lambda, \lambda
'])
\ena

Par exemple, l'alg{\`e}bre de Weil de Lie $\cg$ $\cw (\cg)$ est
d{\'e}finit au moyen des g{\'e}n{\'e}rateurs $\om , \Om$ {\`a} valeur dans
Lie $\cg$, de degr{\'e}s respectifs 1 et 2 avec les {\'e}quations de
structure
\bea
d_W \om &=& \Om - \shalf [\om, \om] \nonumber \\
d_W \Om &=& - [ \om, \Om] \nonumber \\
i_W (\lambda) \om &=& \lambda \; \; \; \; \; i_W (\lambda) \Om =0
\ena

La cohomologie {\'e}quivariante de $M$ est la cohomologie basique de
$\Om^* (M) \otimes \cw (\cg)$ munie de la diff{\'e}rentielle $d_M +
d_W$ et de l'action $i_M (\lambda) + i_W (\lambda)$.

\setcounter{equation}{0}

\annexe{2}{Appendice B}

Nous rappellerons bri{\`e}vement quelques-unes des difficult{\'e}s bien
connues que l'on rencontre en dimension infinie par exemple dans le
cas des th{\'e}ories de jauge {\`a} la Yang Mills. Le point de d{\'e}part,
c'est-{\`a}-dire la forme invariante sur $\ca$
\be
\Theta = \e^{-S_{inv}(a)} \cd a
\ee
n'existe pas pour la m{\^e}me raison que la mesure de Haar sur 
le groupe de jauge n'existe pas. La forme de Ruelle Sullivan a des
chances d'exister, modulo les probl{\`e}mes ultraviolets que,
pr{\'e}cis{\'e}ment, on ne sait pas r{\'e}gler sur l'espace des orbites,
mais on n'en a pas de forme explicite. Le point de d{\'e}part est donc
remplac{\'e} par une classe d'{\'e}quivalence dont l'unicit{\'e} n'est pas
garantie. Il est concevable qu'il en r{\'e}sulte l'unicit{\'e} pour les
observables locales mais pas pour des "observables {\`a} l'infini". Il
ne reste plus qu'{\`a} donner un sens {\`a} des formes diff{\'e}rentielles
de dimension infinies et {\`a} leurs int{\'e}grales qui font
appara{\^\i}tre les comportements ultraviolets usuels qu'on ne sait
ma{\^\i}triser que gr{\^a}ce {\`a} la localit{\'e} dans l'espace des
champs. Quant {\`a} leur existence, bas{\'e}e sur l'existence de partitions
 de l'unit{\'e} sur $\ca / \cg$,
il faut se rappeler que ce dernier espace n'est m{\'e}trique que pour
la topologie $L^2$ \cite{5}. Le probl{\`e}me de r{\'e}concilier des
fixations de jauge g{\'e}om{\'e}triquement licites avec la localit{\'e} de
la th{\'e}orie des champs reste donc ouvert.

\end{document}